\documentclass[pra,twocolumn,superscriptaddress,showpacs]{revtex4}
\usepackage{hyperref}
\usepackage{bbm}
\usepackage{amsmath, amssymb}
\usepackage{color}
\usepackage{graphicx,epsfig}
\usepackage{pifont}
\usepackage{subfigure}
\usepackage{amsmath, amssymb, graphics}

\begin{document}

\title{Markovian evolution of strongly coupled harmonic  oscillators}
\author{Chaitanya Joshi}
\email{chaitanya.heriot@gmail.com}
\affiliation{School of Physics and Astronomy, University of St Andrews, St  Andrews, KY16 9SS, UK }
\author{Patrik  \"Ohberg}
\affiliation{SUPA,  Institute of Photonics and Quantum Sciences,
Heriot-Watt University, Edinburgh, EH14 4AS, UK}
\author{James D. Cresser}
\affiliation{Department of Physics and Astronomy, Macquarie University, 2109 NSW, Australia}
\author{Erika Andersson}
\affiliation{SUPA,  Institute of Photonics and Quantum Sciences,
Heriot-Watt University, Edinburgh, EH14 4AS, UK}

\begin{abstract}
We investigate how to model Markovian evolution of 
coupled harmonic oscillators, each of them interacting with a local environment. 
When the coupling between the oscillators is weak, dissipation may be modeled using local Lindblad terms for each of the oscillators in the master equation, as is commonly done. When the coupling between oscillators is strong, this model may become invalid.
We derive a master equation for two coupled harmonic oscillators which are subject to individual heat baths modeled by a collection of harmonic oscillators, and show that this master equation in general contains non-local Lindblad terms.
We compare the resulting time evolution with that
obtained 
for dissipation through local Lindblad terms for each individual oscillator, and show that the evolution is different in the two cases. In particular, the two descriptions give different predictions for the steady state and for the entanglement between strongly coupled oscillators. This shows that when describing strongly coupled harmonic oscillators, one must take great care in how dissipation is modeled, and that a description using local Lindblad terms may fail. This may be particularly relevant when attempting to generate entangled states of strongly coupled quantum systems.
\end{abstract}

\maketitle
\section{Introduction}

Physical systems will always, to a varying degree, interact with their external environments. This is commonly described using some kind of master equation. Very often approximations are made, resulting in a master equation of so-called Lindblad form. When modelling quantum systems coupled to each other and to their environments, it might therefore be tempting to ``phenomenologically" add decay terms of Lindblad form to account for interaction with an environment, without going via a rigorous derivation of the master equation from first principles. As first noted by Walls, however, one has to be careful when doing this, as naively adding Lindblad terms may give the wrong steady state for strongly coupled quantum systems~\cite{wallsp}.

In this paper we further explore how to describe the system-environment interaction for coupled harmonic oscillators. In particular, we will be especially interested in strongly coupled oscillators, and will consider not just the steady state, but also predictions regarding quantum entanglement. Many different physical systems can be described as coupled harmonic oscillators, including coupled vibrational degrees of freedom of ions in an ion trap \cite{krbr}, quantum fluctuations of mechanical and optical modes around the classical steady state value in an optomechanical cavity  \cite{smans3, agarwal, akram, joshopt}, or coupled nano-sized electromechanical devices arranged in an array \cite{eisbos}. Entanglement properties of such systems are of great current interest in the context of quantum information science and quantum technology. This is one motivation for our work.

The paper is organized as follows. In Section \ref{sec:me} we briefly review the derivation of master equations.
In section \ref{sec:pheno}, a master equation describing the evolution of two coupled oscillators is examined,
for the case when each oscillator undergoes  damping through a local Lindblad term, which could arise as a ``phenomenologically" motivated master equation. Following this,
a Markovian master equation describing the evolution of two strongly coupled oscillators is derived in section \ref{sec:mastersolexct}, assuming that each oscillator interacts with a local bath of harmonic oscillators.  In  section \ref{sec:resltsapp} we compare the  evolution of the two coupled oscillators when dissipation is modeled using the above two approaches. We compare oscillator excitations, fidelity between the two mode states for the different damping models, and the quantum correlations between the two coupled oscillators, as measured by the logarithmic negativity, and show these to be different, in general, in the two cases. The observation made by Walls for the steady state is also confirmed for zero temperature reservoirs. Finally, we conclude the paper with  discussions in section \ref{sec:concld}.

\section{Deriving master equations}
\label{sec:me}

To illustrate how master equations are in general derived, we will be considering two bilinearly 
coupled harmonic oscillators with equal frequency $\omega$ and unit mass, labelled by $a$ and $b$, described by a Hamiltonian
(with $\hbar=1$) 
\begin{multline}
	H_{\text{sys}}=\frac{1}{2}\left(\hat{p}_a^2+\omega^2\hat{x}_a^2\right)+\frac{1}{2}\left(\hat{p}_b^2+\omega^2\hat{x}_b^2\right)+\frac{1}{2}\Omega^2\left(\hat{x}_a\hat{x}_b\right).
\end{multline}
Expressing the position and momentum quadratures in terms of annihilation and creation operators, and setting 
$\kappa=\Omega^2/4\omega$, we can write the Hamiltonian as 
\begin{equation}
\label{eq:sysham}
	\begin{split}
		H_\text{sys}
		=&H_a+H_b+H_\text{coupling}\\
		=&\omega\left(\hat{a}^\dagger\hat{a}
		+\hat{b}^\dagger\hat{b}\right)
		+\kappa\left(\hat{a}+\hat{a}^\dagger\right)\left(\hat{b}+\hat{b}^\dagger\right)
	\end{split}
\end{equation}
where $\hat a$ and $\hat b$ are destruction operators for the individual modes of the two coupled oscillators.

The joint state of the two oscillators and their outside environment evolve under the Hamiltonian
\begin{equation}
H=H_\text{sys}+H_\text{env}+H_\text{int}.
\end{equation}
The environment is often modeled as a collection of harmonic oscillators linearly coupled to the system oscillators through
\begin{equation}
	H_\text{int}=\hat{R}_a(\hat{a}+\hat{a}^\dagger)+\hat{R}_b(\hat{b}+\hat{b}^\dagger),
\end{equation}
where the reservoir operators $\hat{R}_{a,b}$ take the generic form
\begin{equation}
	\hat{R}=\sum_{\Omega}^{}f(\Omega)(\hat{c}^\dagger_\Omega+\hat{c}_\Omega),
\end{equation}
and $\hat c_\Omega$ refer to the modes of the environment which interact with the quantum system with interaction strength  $f(\Omega)$.
It is sometimes possible to derive an exact master equation, 
such as the Hu-Paz-Zhang master equation for the Caldeira-Leggett model, when 
the system is a single harmonic oscillator \cite{HuPazZhang}. 
The case of coupled harmonic oscillators interacting with an arbitrary number of reservoirs has been dealt with by influence functional methods in \cite{martinez}. The results are exact, but forbiddingly complex in the general case. We are working in the limit in which the Born-Markov approximation is appropriate, and a limiting case could in principle be arrived at from the exact results of \cite{martinez} by appropriate choice of spectral density and time scales. For our investigations, however,Êwe have chosen more usual methods, which are sufficient to illustrate the point we wish to make.

First, the rotating wave approximation (RWA) can be made on the coupling between the oscillators.
Second, approximations including the RWA can be made for the interaction between
the oscillators and the environment. The RWA involves dropping quickly oscillating terms from the Hamiltonian.
In commonly encountered physical situations, the oscillators are weakly coupled to each other, so that $\kappa \ll \omega$. If this condition is met, then the coupling 
between the oscillators can be simplified under the rotating wave approximation (RWA) \cite{stev,milburn}. 
The two-mode squeezing terms $\hat{a} \hat{b}$ and $\hat{b}^{\dagger} \hat{a}^{\dagger}$ oscillate, in the interaction picture, on a time scale $\sim$ (system transition frequencies)$^{-1}$, i.e., $\omega^{-1}$, which, if the RWA can be made, is much faster than the time scale $\kappa^{-1}$ of the coupling between the oscillators. The two-mode squeezing terms can then be dropped from the Hamiltonian. The RWA Hamiltonian then conserves the number of excitations and can be straightforwardly solved using a simple rotational transformation to the center of mass and the relative modes of the two oscillators.

Similar reasonings hold when applying the RWA on the interaction between a system and its environment. The ``system" could also consist of several subsystems, each one in their independent environments, or in a common environment. In any case, properly applying the RWA should again involve dropping quickly oscillating terms. 

For the interaction of the system with the environment, the theory of open quantum systems provides a suite of approximations, including the Born-Markov and secular approximations (the BMS approximations), that have been found to be widely valid. The secular approximation amounts to
excluding rapidly oscillating terms arising in
the master equation. Alternatively, the usual RWA can be introduced directly at the level of the Hamiltonian. Either way, this approximation is essential -- the Born-Markov approximations are not enough -- in yielding a master equation for the system density operator $\rho$ that is of Lindblad form (ignoring energy shifts),
\begin{equation}
	\dot{\rho}=-i\left[H_\text{sys},\rho\right]+\sum_{s}^{}\gamma_s\mathcal{L}_{s}[\rho],
\end{equation}
where $\mathcal{L}_{s}\rho=\hat s\rho\,\hat s^\dagger-\frac{1}{2}\left\{\hat s^\dagger \hat s,\rho\right\}$ and where $\hat s,\hat s^\dagger$ are harmonic oscillator annihilation and
creation operators whose detailed form depends on 
not just the form of the system-reservoir interaction, 
but also on the coupling between the systems, in our case the two oscillators. 

For a system of coupled oscillators, each independently interacting with  separate heat baths, one might expect that the form of the damping for each coupled oscillator is independent of the strength of the coupling between them. Therefore, a naive approach, which can also be referred to as a \emph{local} damping model, is simply to add local Lindblad terms for each oscillator~\cite{smans3, akram, joshopt, geradolpa, josh}. That is, the Lindblad operators $\hat s,\hat s^\dagger$ would be given by $\hat a,\hat a^\dagger, \hat b, \hat b^\dagger$ to yield
\begin{multline}\label{mastereqntwoosc3}
	\dot{\rho}=-i\left[H_\text{sys},\rho\right]\\+\sum_{s=a,b}^{}\left\{\Gamma_s\left(\bar{n}_s+1\right)\mathcal{L}_s[\rho]+\Gamma_s\bar{n}_s\mathcal{L}_{s^\dagger}[\rho]\right\}
\end{multline}
where $\Gamma_{s}$ is the decay rate of oscillator $s$, which is coupled to a heat bath with average thermal occupancy $\bar{n}_{s}$. 
This approach is certainly valid if the coupling between the oscillators is
weak. 
As was first noted by Walls~\cite{wallsp}, however, it may also fail.
Walls investigated the case of  two coupled bosonic  modes with inter-mode coupling $H_\text{coupling}=\kappa(\hat{a}^\dagger\hat{b}+\hat{b}^\dagger \hat{a})$. He pointed out that for inter-mode coupling $\kappa \sim \omega$, modeling dissipation through local damping of each individual mode may become questionable. The steady state density operator for the system ought to be the canonical density operator 
\begin{equation}\label{canonical}
	\rho(\infty)\sim \exp\left(-H_\text{sys}/k_BT\right)
\end{equation}
in the limit of weak coupling of the system(s) to the reservoir(s) at a temperature $T$. This is, of course, the expected steady state density operator for the system in thermal equilibrium with the reservoir at temperature $T$.

 If, instead, the damping of Lindblad form is added naively, with $\hat s$ equal to the individual system mode operators $\hat a$ and $\hat b$,
then $\hat{H}_a+\hat{H}_b$ appears in the exponent. While this might be an acceptable approximation for weak inter-system coupling
($\kappa\ll\omega$), in the limit of strongly coupled oscillators it is not a valid result. 

When the coupling between oscillators is strong, it is necessary 
to derive the master equation in a way that fully accounts for the coupling between the oscillators. The Lindblad operators $\hat s,\hat s^\dagger$ are then the energy eigenoperators \cite{gardiner} for the composite system, or equivalently the normal modes of the coupled harmonic oscillators, instead of the operators  $\hat a,\hat a^\dagger, \hat b, \hat b^\dagger$. Working with the eigenoperators is of course a general principle when it comes to dealing with damping of any composite quantum system -- it is not confined to coupled harmonic oscillators. For instance, the ``correct'' damping of the Jaynes-Cummings model resulting in an equilibrium steady state entails an eigenoperator approach \cite{jim}. 
But in doing so, perhaps somewhat counter-intuitively, the Lindblad terms in the resulting master equations are then in a sense \emph{non-local}. The resulting time evolution then differs from the one obtained using a master equation with local Lindblad terms. In other words, when the oscillators are strongly coupled to each other, a description using local Lindblad terms may fail, and it is this regime we are interested in. In particular in this regime of strong coupling we find, for example, that we would not expect to be able to make the RWA in the oscillator-oscillator coupling in the Hamiltonian \eqref{eq:sysham}.  In order to study in greater detail the contributions of the non-RWA terms, we will work, instead of with the Hamiltonian \eqref{eq:sysham}, but rather with a generalized Hamiltonian for the two oscillators,
\begin{equation}\label{eqn2ham}
	H_\text{sys}=\omega (\hat{a}^{\dagger} \hat{a}+\hat{b}^{\dagger} \hat{b})
	+\kappa (\hat{a}^{\dagger}\hat{b}+\hat{b}^{\dagger} \hat{a})
	+\lambda (\hat{a}\hat{b}+\hat{b}^{\dagger}\hat{a}^{\dagger}).
\end{equation}
From this Hamiltonian, \eqref{eq:sysham} can be regained in the case of $\lambda=\kappa$, and the RWA form for $\lambda=0$. Hamiltonian \eqref{eqn2ham} however also includes e.g. the case of a squeezing interaction, which is obtained for $\kappa =0$. Furthermore, we are here interested in the evolution of coupled harmonic oscillators initially prepared in Gaussian states. The generalized Hamiltonian \eqref{eqn2ham} is still quadratic in position and momentum coordinates, and hence the initial Gaussian nature of the state is preserved.

\section{Local Lindblad-type dissipation} 
\label{sec:pheno}

As stated in the Introduction, there is inevitable coupling between the system of interest and the environment. In this section, we shall consider the case where dissipative dynamics is modeled by adding local Lindblad operators for each individual oscillator.  As we will show, adding such local Lindblad terms to the master equation must be carefully justified, and may in fact lead to incorrect dynamics  if the oscillators are strongly coupled to each other.
For now, we nevertheless assume that the time evolution of the system of two coupled harmonic oscillators in the  Born-Markov approximation is  described by the Lindblad-type master equation \eqref{mastereqntwoosc3}, repeated here,
\begin{equation*}
\frac{\partial \rho }{\partial t}=-i[H_\text{sys},\rho]+
\sum_{s=a,b}
\{[\Gamma_s(\bar{n}_{s}+1)\mathcal L_s(\rho)+\Gamma_s\bar{n}_s\mathcal L_{s^\dagger}(\rho)]\}.
\end{equation*}

If the two coupled oscillators interact with an environment which itself can be described by a Gaussian state, with an interaction Hamiltonian that contains  terms at most quadratic in annihilation and creation operators, then the two coupled oscillators maintain their initial Gaussian character during the resulting dissipative evolution. One way to  solve the master equation \eqref{mastereqntwoosc3} is to exploit this Gaussian character, by rewriting the master equation in terms of a partial differential equation for the two-mode quantum characteristic function. We define a normal-ordered characteristic function as $\chi(\kappa_{a},\kappa_{a}^{*},\eta_{b},\eta_{b}^{*},{\it t})=\langle e^{\kappa_{a}\hat{a}^{\dagger}}e^{-\kappa_{a}^{*}\hat{a}}e^{\eta_{b}\hat{b}^{\dagger}}e^{-\eta_{b}^{*}\hat{b}}\rangle$ and make a Gaussian ansatz for the time-evolved  characteristic function, $\chi(\kappa_{a},\kappa_{a}^{*},\eta_{b},\eta_{b}^{*},{\it t}) =\exp\left[-z^{T}{\mbox{\bf L}}(t)\,z+i z^{T}h(t)\right]$.
 Here ${\mbox{\bf L}}(t)$ is a time-dependent 4$\times$4 symmetric matrix, $h(t)$ is a 4$\times$1 time-dependent vector and $z^{T}=( \kappa_{a},\kappa_{a}^{*},\eta_{b},\eta_{b}^{*})$.  The corresponding partial differential equation for $\chi(\kappa_{a},\kappa_{a}^{*},\eta_{b},\eta_{b}^{*},{\it t})$ then becomes \cite{stev}
\begin{equation}\label{chievnay}
\frac{\partial }{\partial t} \chi
=z^{T}  {\mbox{\bf M}} z \chi+z^{T}  {\mbox{\bf N}} \nabla \chi,
\end{equation}
where
$\nabla= (\frac{\partial}{ \partial \kappa_{a}}, \frac{\partial }{ \partial \kappa_{a}^{*}}, \frac{\partial}{ \partial \eta_{b}}, \frac{\partial }{ \partial \eta_{b}^{*}})^{T}$
and
\begin{eqnarray}
{\mbox{\bf N}}&=&\left( \begin{array}{cccc}
   i\omega-\Gamma_{a} & 0 & i \kappa &-i\lambda \\
     0 & -i\omega-\Gamma_{a} & i \lambda &-i \kappa  \\
      i \kappa  & -i\lambda & i\omega-\Gamma_{b}&0 \\
       i\lambda  & -i \kappa  &0 &-i\omega-\Gamma_{b}\\
  \end{array} \right ),\\
{\mbox{\bf M}}&=&\left( \begin{array}{cccc}
  0 & -\Gamma_{a} \bar{n}_{a} & i \lambda/2&0 \\
     -\Gamma_{a} \bar{n}_{a} & 0 & 0&-i \lambda/2\\
      i \lambda/2 & 0 & 0&-\Gamma_{b}\bar{n}_{b} \\
       0 & -i \lambda/2 &-\Gamma_{b} \bar{n}_{b} &0 \\
  \end{array} \right ).
  \end{eqnarray}
Using the Gaussian ansatz for the quantum characteristic function $\chi(\kappa_{a},\kappa_{a}^{*},\eta_{b},\eta_{b}^{*},{\it t})$  it easily follows that 
\begin{eqnarray}
\frac{\partial \chi }{ \partial t}&=&-z^{\rm T} \frac{d {\mbox{\bf L}}}{d t} z \chi+i z^{\rm T} \frac{d h}{d t} \chi \label{diff1}\\ \nabla \chi &=&-{\rm 2} {\mbox{\bf L}} z \chi +i h \chi.\label{diff2}
\end{eqnarray}
Using \eqref{diff1} and \eqref{diff2}, the partial differential equation \eqref{chievnay} for $\chi$ becomes
\begin{equation}\label{difchmon}
-z^{\rm T} \frac{d {\mbox{\bf L}}}{d t} z+i z^{\rm T} \frac{d h}{d t} \chi = z^{\rm T} {\mbox{\bf M}} z \chi-{\rm 2}z^{\rm T}{\mbox{\bf N}}{\mbox {\bf L}} z \chi +i z^{\rm T} {\mbox{\bf N}} h \chi.
\end{equation}
Recalling that $\mbox{\bf {L}}(t)$ is symmetric, we can write
  \begin{equation}
  {\mbox{\bf L}}(t)=\left( \begin{array}{cc}
 {\mbox{\bf P}}(t)& {\mbox{\bf Q}}(t)\\
    {\mbox{\bf Q}}(t)^{T}& {\mbox{\bf R}}(t)\\
  \end{array} \right),
  \end{equation}
 where ${\mbox{\bf P}}(t)$ and ${\mbox{\bf R}}(t)$ are 2$\times$2 symmetric matrices. Taking the symmetric part of  \eqref{difchmon} results in two matrix differential equations
\begin{align}\label{nayitue}
\frac{d {\mbox{\bf L}}(t)}{d t}+ {\mbox{\bf M}}=&{\mbox{\bf N}}{\mbox{\bf L}}+{\mbox{\bf L}} {\mbox{\bf N}}^{\rm T} \\
\frac{d h}{d t}=& \mbox{\bf N} h.
  \end{align}
Thus solving the master equation \eqref{mastereqntwoosc3} reduces to solving two coupled matrix differential equations. From  the quantum characteristic function the complete statistical description of the corresponding state can be obtained. With the quantum  characteristic function one can therefore also obtain the expectation values of quantum mechanical observables, e.g.
\begin{multline*}
	\langle \hat{a}^{\dagger m}(t){\it \hat{b}^{\dagger n}}(t) \rangle\\=\it {\left(\frac{\partial }{\partial \kappa_{a}}\right)^{m}\left(\frac{\partial }{\partial \kappa_{b}}\right)^{n}\chi(\kappa_{a},\kappa_{a}^{*},\eta_{b},\eta_{b}^{*},{\it t})}|_{\kappa_{a},\kappa_{a}^{*},\eta_{b},\eta_{b}^{*}=\rm{0}}.
\end{multline*} 
In the next section  we shall  derive a master equation describing the dynamics of two strongly coupled harmonic oscillators where their individual environments are modeled as collections of harmonic oscillators. By numerically solving both the master equation in \eqref{mastereqntwoosc3}, describing local Lindblad-type dissipation, and the master equation derived in the next section,  we shall then, in section \ref{sec:resltsapp}, compare the dissipative evolution of the two strongly coupled harmonic oscillators in these two cases.

\section{Bath-induced dissipation}
\label{sec:mastersolexct}
A master equation of standard Lindblad form guarantees the positivity of the time-evolved density matrix. It may seem justified to simply add local  Lindblad terms acting on the individual coupled oscillators $a$ and $b$, also when they are strongly coupled to each other, as was done in the previous section. As we will show next, however, this is fraught with pitfalls. In what follows we shall derive a Markovian master equation for two strongly coupled harmonic oscillators which are harmonically coupled to their local heat baths.The result is a master equation  of Lindblad form, but the Lindblad superoperators $\mathcal L$ do not act locally on each individual oscillator.

The oscillators are as before labelled $a$ and $b$ and their coupled dynamics is governed by the  Hamiltonian \eqref{eqn2ham}. We consider a scenario where the two oscillators are irreversibly coupled to local heat baths, each of which is modeled as a collection of many harmonic oscillators. The Hamiltonian corresponding to the two independent  local heat baths is given by 
\begin{equation}
H_{\rm env}= \sum_{\Omega}\Omega \hat{c}^{\dagger}_{\Omega} \hat{c}_{\Omega} +\sum_{\Omega'} \Omega'  \hat{d}^{\dagger}_{\Omega'} \hat{d}_{\Omega'} ,
\end{equation}
where $\hat{c}_{\Omega}$ and $\hat{d}_{\Omega'}$ represent the destruction operators for the bosonic modes of the local heat baths for oscillators $a$ and $b$, respectively.  Assuming a bilinear coupling between the position quadratures of each oscillator and the modes of their local heat baths,  the system-environment interaction takes the form

\begin{equation}\label{hamilappnd3}
 H_{\rm int}= \sum_{\Omega} \zeta_{\Omega} (\hat{c}^{\dagger}_{\Omega}+\hat{c}_{\Omega}) (\hat{a}+\hat{a}^{\dagger})+ \sum_{\Omega'} \eta_{\Omega'} (\hat{d}^{\dagger}_{\Omega'}+\hat{d}_{\Omega'}) (\hat{b}+\hat{b}^{\dagger}),
\end{equation}
where $\zeta_{\Omega}$ and $\eta_{\Omega'}$ are the coupling strengths between each individual oscillator and modes of the corresponding environment. The two coupled oscillators undergo unitary evolution described by the Hamiltonian
\begin{equation} \label{mastcrctapprch}
\hat{H}=\hat{H}_{\rm sys}+\hat{H}_{\rm env}+\hat{H}_{\rm int}.
\end{equation}
Using  the Hamiltonian \eqref{mastcrctapprch}, a master equation describing the dissipative evolution of the two coupled oscillators will now be derived.

\subsection{Derivation of the coupled oscillator master equation}
\label{sec:appddx}

In order to derive a master equation for the two coupled harmonic oscillators we shall first diagonalize the Hamiltonian \eqref{eqn2ham}  by defining the center of mass and relative modes,
\begin{eqnarray}\label{efab}
\hat{e}&=&\frac{\hat{a}+\hat{b}}{\sqrt{2}}\\
\hat{f}&=&\frac{\hat{a}-\hat{b}}{\sqrt{2}}.
\end{eqnarray}
The Hamiltonian \eqref{eqn2ham} now becomes 
\begin{eqnarray}\label{hamilsqezng}
H_{\rm sys}&=& \omega (\hat{e}^{\dagger}\hat{e}+\hat{f}^{\dagger} \hat{f}) +\frac{\lambda}{2} (\hat{e}^{2}+\hat{e}^{ \dagger 2}-\hat{f}^{2}-\hat{f}^{ \dagger 2})\nonumber\\&&+\frac{\kappa}{2}(\hat{e}\hat{e}^{\dagger}+\hat{e}^{\dagger} \hat{e} -\hat{f}\hat{f}^{\dagger}-\hat{f}^{\dagger} \hat{f}), 
\end{eqnarray}
which can be diagonalized using a Bogoliubov transformation
\begin{eqnarray}\label{secappndx3}
\left( \begin{array}{c} 
 {\hat{e}} \\
 {\hat{e}^{\dagger}}  
\end{array} 
\right)&=& \left( \begin{array}{cc} 
 {\alpha_{1}} & {-\beta_{1}}\\
 {-\beta_{1}} & {\alpha_{1}} 
\end{array} 
\right)\left( \begin{array}{c} 
\hat l \\
 \hat l^\dagger   
\end{array}\right),\\ 
\left( \begin{array}{c} 
 {\hat{f}} \\
 {\hat{f}^{\dagger}}  
\end{array} 
\right)&=&\left( \begin{array}{cc} 
 {\alpha_{2}} & {-\beta_{2}}\\
 {-\beta_{2}} & {\alpha_{2}} 
\end{array} 
\right)\left( \begin{array}{c} 
 \hat m \\
 \hat m^\dagger 
\end{array}\right).\\ \nonumber
\end{eqnarray}
The Hamiltonian \eqref{hamilsqezng}  then takes the simplified form
\begin{equation}
H_{\rm sys}= (\alpha_{11}+\alpha_{22}) \hat{l}^{\dagger} \hat{l}+ (\beta_{11}+\beta_{22})\hat{m}^{\dagger} \hat{m},
\end{equation}
with 
\begin{eqnarray}
\alpha_{11}=\frac{(2 \omega +\kappa) \alpha_{1}^{2}-2 \lambda \alpha_{1}\beta_{1}+\kappa \beta_{1}^{2}}{2}\\
\alpha_{22}=\frac{(2 \omega +\kappa) \beta_{1}^{2}-2 \lambda \alpha_{1}\beta_{1}+\kappa \alpha_{1}^{2}}{2}\\
\beta_{11}=\frac{(2 \omega -\kappa) \alpha_{2}^{2}+2 \lambda \alpha_{2}\beta_{2}-\kappa \beta_{2}^{2}}{2}\\
\beta_{22}=\frac{(2 \omega -\kappa) \beta_{2}^{2}+2 \lambda \alpha_{2}\beta_{2}-\kappa \alpha_{2}^{2}}{2},
\end{eqnarray}
where $\alpha_{i}$ and $\beta_{i}$ are of the form 
\begin{eqnarray}
\alpha_{1}^{2}&=&\frac{1}{2}+\frac{1}{2}\frac{\kappa+\omega}{\sqrt{(\kappa+\omega)^{2}-\lambda^{2}}}\\
\beta_{1}^{2}&=&-\frac{1}{2}+\frac{1}{2}\frac{\kappa+\omega}{\sqrt{(\kappa+\omega)^{2}-\lambda^{2}}}\\
\alpha_{2}^{2}&=&\frac{1}{2}+\frac{1}{2}\frac{-\kappa+\omega}{\sqrt{(-\kappa+\omega)^{2}-\lambda^{2}}}\\
\beta_{2}^{2}&=&-\frac{1}{2}+\frac{1}{2}\frac{-\kappa+\omega}{\sqrt{(-\kappa+\omega)^{2}-\lambda^{2}}}.
\end{eqnarray}
\begin{widetext}
For the coupled oscillators to maintain their oscillatory behaviour,  $\lambda < |\omega-\kappa|$ is required. Thus the free evolution of the two coupled oscillators and their local environment is given by
\begin{eqnarray}
H_{\rm sys}+H_{\rm env}= 
(\alpha_{11}+\alpha_{22}) \hat{l}^{\dagger} \hat{l}+ (\beta_{11}+\beta_{22})\hat{m}^{\dagger} \hat{m}
+\nonumber
\sum_{\Omega}\Omega \hat{c}^{\dagger}_{\Omega} \hat{c}_{\Omega} +\sum_{\Omega'} \Omega'  \hat{d}^{\dagger}_{\Omega'} \hat{d}_{\Omega'}.
\end{eqnarray}
Re-expressing the bare modes $a$ and $b$ in terms of  $\hat{l}$ and $\hat{m}$, the Hamiltonian \eqref{mastcrctapprch} in the interaction picture with $H_0=H_{\rm sys}+H_{\rm env}$ becomes
\begin{eqnarray}\label{eqnhamapp3sol}
H_{I}(t)&=& \sum_{\Omega,\Omega'}[\zeta_{\Omega}(\hat{c}_{I}+\hat{c}_{I}^{\dagger}) (\alpha_{1}\hat{l}_{I}-\beta_{1}\hat{l}_{I}^{\dagger}+\alpha_{1}\hat{l}_{I}^{\dagger}-\beta_{1}\hat{l}_{I}+\alpha_{2}\hat{m}_{I}-\beta_{2}\hat{m}_{I}^{\dagger}+\alpha_{2}\hat{m}_{I}^{\dagger}-\beta_{2}\hat{m}_{I}) \nonumber\\&&
+\eta_{\Omega'}(\hat{d}_{I}+\hat{d}_{I}^{\dagger}) (\alpha_{1}\hat{l}_{I}-\beta_{1}\hat{l}_{I}^{\dagger}+\alpha_{1}\hat{l}_{I}^{\dagger}-\beta_{1}\hat{l}_{I}-\alpha_{2}\hat{m}_{I}+\beta_{2}\hat{m}_{I}^{\dagger}-\alpha_{2}\hat{m}_{I}^{\dagger}+\beta_{2}\hat{m}_{I})],
\end{eqnarray}
\end{widetext}
where  $H_{I}(t)=e^{-iH_{0}t}H_{{\rm int}}e^{i H_{0} t}$, $\hat{l}_{I}$=$\hat{l} ~e^{-i (\alpha_{11}+\alpha_{22})t}$,  $\hat{m}_{I} =\hat{m} ~e^{-i (\beta_{11}+\beta_{22})t}$, $\hat{c}_{I}=\hat{c}_{\Omega}~ e^{-i\Omega t}$, $\hat{d}_{I}=\hat{d}_{\Omega'}~ e^{-i\Omega' t}$ and a factor of $1/ \sqrt{2}$ has been absorbed into the definition of $\zeta_{\Omega}$ and $\eta_{\Omega'}$. 

If the system-reservoir coupling is weak we can simplify the interaction Hamiltonian \eqref{eqnhamapp3sol} using the rotating wave approximation (RWA). Invoking the RWA essentially amounts to dropping the fast oscillating terms proportional to $\hat{c}_{I}\hat{l}_{I}, \hat{c}_{I}\hat{m}_{I},\hat{d}_{I}\hat{l}_{I},\hat{d}_{I}\hat{m}_{I}$ and their Hermitian conjugates from the Hamiltonian \eqref{eqnhamapp3sol}, which results in
 \begin{equation}\label{eqnhamapp3appslnew}
 H_{I}(t)= \hat{l}_{I} \hat{F}^{\dagger}(t)+\hat{m}_{I} \hat{Q}^{\dagger}(t)+h.c.,
 \end{equation}
where the  {\rm noise} operators are given by  
\begin{eqnarray}
\hat{F}^{\dagger}(t)= \sum_{\Omega,\Omega'} (\alpha_{1}- \beta_{1})(\zeta_{\Omega} \hat{c}_{\Omega}^{\dagger} e^{i \Omega t}+\eta_{\Omega'}  \hat{d}_{\Omega'}^{\dagger}e^{i \Omega' t}),\\
\hat{Q}^{\dagger}(t)= \sum_{\Omega,\Omega'} (\alpha_{2}- \beta_{2})(\zeta_{\Omega} \hat{c}_{\Omega}^{\dagger} e^{i \Omega t}-\eta_{\Omega'}  \hat{d}_{\Omega'}^{\dagger}e^{i \Omega' t}).
\end{eqnarray}
Using the RWA and in the interaction picture with $H_0=H_{\rm sys}+H_{\rm env}$, the joint state of the oscillators and their local environments, represented by the total density matrix $\rho_{I}$, evolves according to
 \begin{equation}\label{totdenstymat}
 \dot{\rho_{I}}(t)=-{i}[H_{I}(t), \rho_{I}(t)],
 \end{equation}
 where $H_{I}(t)$ is given by Eq. \eqref{eqnhamapp3appslnew}. We assume that at $t=0$ the joint state of the system and environments is factorizable so that $\rho_{I}(t=0)= \rho_{\rm e}(0) \otimes  \rho_{\rm sys}(0)$ where $\rho_{\rm e}(0)$ is the joint initial state of the two local baths and $\rho_{\rm sys}(0)$ is the density matrix of the  two coupled harmonic oscillators.

The evolution of the density matrix $\rho_{\rm sys}$ representing the state of the two oscillators  is  given by
 \begin{equation}\label{denstymatrxappds3}
 \dot{\rho}_{\rm sys}(t)={\rm Tr_{e}}\dot{\rho}_{I}(t)=-i{\rm Tr_{e}}[H_{I}(t), \rho_{I}(t)],
 \end{equation}
 where ${\rm Tr_{e}}$ denotes the trace over the environmental degrees of freedom. If we also assume that the state of the environment for each oscillator remains unaffected as a result of the coupling, then the joint state of the system evolves as  $\rho_{I}(t)= \rho_{\rm e}(0) \otimes \rho_{\rm sys}(t)$.  Formally integrating \eqref{totdenstymat} gives 
 \begin{equation}
 \rho_{I}(t)=\rho_{I}(0)-i \int_{0}^t[H_{I}(t'),\rho_{I}(t')] dt',
 \end{equation}
which when substituted in \eqref{denstymatrxappds3} gives an integro-differential equation for the state of the oscillators,
\begin{widetext}
 \begin{eqnarray}\label{mstappnd3intgrl}
 \dot{\rho}_{\rm sys}(t)&=&-i {\rm Tr_{e}}[ H_{I}(t),\rho_{I}(0)]-
 \int_{0}^t {\rm Tr_{e}}[H_{I}(t),[H_{I}(t'),\rho_{I}(t')]] dt'.
 \end{eqnarray}
For an environment in thermal equilibrium the first term in \eqref{mstappnd3intgrl} is identically zero. Using \eqref{eqnhamapp3appslnew} the above integro-differential equation takes the form
	\begin{equation}\label{densmatapp}
		\dot{\rho}_{\rm sys}(t)=- \int_{0}^{t} {\rm Tr_{e}}\big[ \hat{l}_{I}
		\hat{F}^{\dagger}(t)+\hat{m}_{I} \hat{Q}^{\dagger}(t)+h.c.,[ \hat{l}_{I}
		\hat{F}^{\dagger}(t')+\hat{m}_{I} \hat{Q}^{\dagger}(t')+h.c., \rho_{\rm e}(0) 
		\otimes \rho_{\rm sys}(t')]] dt'.
  \end{equation}
\end{widetext}
Equation \eqref{densmatapp} can be rearranged as 
\begin{eqnarray}
   \dot{\rho}_{\rm sys}(t)&=&-\int_{0}^{t}{\rm Tr_{e}}[H_{I}(t)H_{I}(t')\rho_{\rm e}(0) \otimes \rho_{\rm sys}(t')\nonumber\\
 && -H_{I}(t)\rho_{\rm e}(0) \otimes \rho_{\rm sys}(t') H_{I}(t')
 \nonumber \\ 
 &&-H_{I}(t')\rho_{\rm e}(0) \otimes \rho_{\rm sys}(t') H_{I}(t)\nonumber \\ 
 && +\rho_{\rm e}(0) \otimes \rho_{\rm sys}(t') H_{I}(t')H_{I}(t)]dt'.
\end{eqnarray}
For environments in thermal equilibrium with flat spectral densities such that $\zeta_{\Omega}=\zeta$ and $\eta_{\Omega'}=\eta$, together with the Markov approximation, one obtains
\begin{eqnarray}
  \sum_{\Omega}\zeta_{\Omega}^{2} e^{i \Omega (t'-t)}&=&\zeta^{2} 2 \pi \delta(t'-t),\\
    \sum_{\Omega'}\eta_{\Omega'}^{2} e^{i \Omega' (t'-t)}&=&\eta^{2}2 \pi \delta(t'-t).
 \end{eqnarray}
One can easily verify that in the case of symmetric coupling of each oscillator to its own environment at zero temperature such that $\pi \zeta^{2}=\pi \eta^{2}=\Gamma$, one obtains the following master equation in the Schr\"odinger picture,
\begin{eqnarray}\label{disspfullappdxnewold}
   \dot{\rho}(t)&=&-i [(\alpha_{11}+\alpha_{22}) \hat{l}^{\dagger}\hat{l}+ (\beta_{11}+\beta_{22})\hat{m}^{\dagger} \hat{m},\rho(t)] \nonumber\\
&&+\langle \hat{F} \hat{F}^{\dagger} \rangle[2 \hat{l} \rho(t) \hat{l}^{\dagger}-\hat{l}^{\dagger}\hat{l}\rho(t)-\rho(t) \hat{l}^{\dagger}\hat{l}]\\
&&+ \langle \hat{Q} \hat{Q}^{\dagger} \rangle[2 \hat{m} \rho(t) \hat{m}^{\dagger}-\hat{m}^{\dagger}\hat{m}\rho(t) -\rho(t) \hat{m}^{\dagger}\hat{m}],\nonumber
\end{eqnarray} 
where $\rho(t)=e^{-iH_{\rm sys} t}\rho_{\rm sys}(t)e^{iH_{\rm sys} t}$ is the density matrix representing the state of the two coupled oscillators in the Schr\"odinger picture and the only non-zero two-time noise correlation functions are of the form
\begin{eqnarray}
   \langle \hat{F}\hat{F}^{\dagger}  \rangle&=&2 \Gamma(\alpha_{1}-\beta_{1})^{2}\\    
    \langle \hat{Q}\hat{Q}^{\dagger}  \rangle&=&2 \Gamma(\alpha_{2}-\beta_{2})^{2} .
     \end{eqnarray}
Reverting back to the bare modes $a$ and $b$, the master equation \eqref{disspfullappdxnewold} takes the form 
 \begin{eqnarray}\label{disspfullappdxnewoldnayi}
 \dot{\rho}(t)&=&-i [H_{\rm sys},\rho(t)] \nonumber\\ &&+
 \Gamma_{1}[2 \hat{a} \rho(t) \hat{a}^{\dagger}-\hat{a}^{\dagger}\hat{a}\rho(t)-\rho(t) \hat{a}^{\dagger}\hat{a}] \nonumber\\ &&+
  \Gamma_{1} [2 \hat{b} \rho(t) \hat{b}^{\dagger}-\hat{b}^{\dagger}\hat{b}\rho(t)-\rho(t) \hat{b}^{\dagger}\hat{b}]\nonumber\\ &&+ 
   \Gamma_{2} [2 \hat{a}^{\dagger} \rho(t) \hat{a}-\hat{a}\hat{a}^{\dagger} \rho(t)-\rho(t) \hat{a}\hat{a}^{\dagger}] \nonumber\\ &&+
    \Gamma_{2} [2 \hat{b}^{\dagger} \rho(t) \hat{b}-\hat{b}\hat{b}^{\dagger} \rho(t)-\rho(t) \hat{b}\hat{b}^{\dagger}] \nonumber\\ &&+ 
   \Gamma_{3}[2 \hat{a} \rho(t) \hat{a}-\hat{a} \hat{a}\rho(t)-\rho(t) \hat{a} \hat{a}]\nonumber\\ &&+ 
   \Gamma_{3} [2 \hat{a}^{\dagger} \rho(t) \hat{a}^{\dagger}-\hat{a}^{\dagger}\hat{a}^{\dagger} \rho(t)-\rho(t) \hat{a}^{\dagger}\hat{a}^{\dagger}] \nonumber\\ &&+ 
   \Gamma_{3}[2 \hat{b} \rho(t) \hat{b}-\hat{b} \hat{b}\rho(t)-\rho(t) \hat{b} \hat{b}]\nonumber\\ &&+
   \Gamma_{3} [2 \hat{b}^{\dagger} \rho(t) \hat{b}^{\dagger}-\hat{b}^{\dagger}\hat{b}^{\dagger} \rho(t)-\rho(t) \hat{b}^{\dagger}\hat{b}^{\dagger}] \nonumber\\ &&+
   \Gamma_{4} [2 \hat{a}\rho(t) \hat{b}^{\dagger}-\hat{b}^{\dagger}\hat{a} \rho(t)-\rho(t)\hat{b}^{\dagger} \hat{a}] \nonumber\\&&+
   \Gamma_{4} [2 \hat{b}\rho(t) \hat{a}^{\dagger}-\hat{a}^{\dagger}\hat{b} \rho(t)-\rho(t) \hat{a}^{\dagger}\hat{b}] \nonumber\\ &&+
     \Gamma_{5} [2 \hat{b}^{\dagger} \rho(t) \hat{a}-\hat{a}\hat{b}^{\dagger}\rho(t)-\rho(t) \hat{a}\hat{b}^{\dagger}] \nonumber\\ &&+
       \Gamma_{5} [2 \hat{a}^{\dagger}\rho(t) \hat{b}-\hat{b}\hat{a}^{\dagger} \rho(t)-\rho(t)\hat{b}\hat{a}^{\dagger}] \nonumber\\ &&+
   \Gamma_{6}[2 \hat{b} \rho(t) \hat{a}-\hat{a} \hat{b}\rho(t)-\rho(t) \hat{a} \hat{b}]  \nonumber\\ &&+
   \Gamma_{6}[2 \hat{a} \rho(t) \hat{b}-\hat{a} \hat{b}\rho(t)-\rho(t) \hat{a} \hat{b}]  \nonumber\\ &&+
   \Gamma_{6} [2 \hat{b}^{\dagger} \rho(t) \hat{a}^{\dagger}-\hat{a}^{\dagger}\hat{b}^{\dagger} \rho(t)-\rho(t) \hat{a}^{\dagger}\hat{b}^{\dagger}] \nonumber\\ &&+
   \Gamma_{6} [2 \hat{a}^{\dagger} \rho(t) \hat{b}^{\dagger}-\hat{b}^{\dagger}\hat{a}^{\dagger} \rho(t)-\rho(t) \hat{a}^{\dagger}\hat{b}^{\dagger}], 
  \end{eqnarray} 
 where the $\Gamma_{i}$ are given by 
 \begin{eqnarray}
 \Gamma_{1}&=&  (\langle \hat{F}\hat{F}^{\dagger}  \rangle \alpha_{1}^{2}+\langle \hat{Q}\hat{Q}^{\dagger}  \rangle \alpha_{2}^{2})/2\\ 
 \Gamma_{2}&=& ( \langle \hat{F}\hat{F}^{\dagger}  \rangle \beta_{1}^{2}+\langle \hat{Q}\hat{Q}^{\dagger}  \rangle \beta_{2}^{2})/2 \\ 
 \Gamma_{3}&=& (\langle \hat{F}\hat{F}^{\dagger}  \rangle \alpha_{1} \beta_{1}+\langle \hat{Q}\hat{Q}^{\dagger}  \rangle \alpha_{2} \beta_{2})/2\\ 
  \Gamma_{4}&=& ( \langle \hat{F}\hat{F}^{\dagger}  \rangle \alpha_{1}^{2}-\langle \hat{Q}\hat{Q}^{\dagger}  \rangle \alpha_{2}^{2} )/2\\ 
 \Gamma_{5}&=& (\langle \hat{F}\hat{F}^{\dagger}  \rangle \beta_{1}^{2}-\langle \hat{Q}\hat{Q}^{\dagger}  \rangle \beta_{2}^{2} )/2\\ 
 \Gamma_{6}&=& ( \langle \hat{F}\hat{F}^{\dagger}  \rangle \alpha_{1} \beta_{1}-\langle \hat{Q}\hat{Q}^{\dagger}  \rangle \alpha_{2} \beta_{2})/2. \\&&\nonumber
 \end{eqnarray}
Equation \eqref{disspfullappdxnewoldnayi} is the final form of the master equation describing the dynamics of two  coupled harmonic oscillators interacting with independent zero temperature baths with flat spectral densities. 
 
It is worth comparing the form of the master equation \eqref{disspfullappdxnewoldnayi}, obtained in the RWA limit when $\lambda={\rm 0}$ with the master equation \eqref{mastereqntwoosc3} in the corresponding case. It is easy to check that when $\lambda$=0,  $\alpha_{1}=\alpha_{2}=$1 and $\beta_{1}=\beta_{2}=$0, 
the master equation \eqref{disspfullappdxnewoldnayi} reduces to the form 
  \begin{eqnarray}\label{mastreqnewjuly}
 \dot{\rho}(t)&=&-i [H_{\rm sys},\rho(t)] \nonumber\\ &&+
 \Gamma_{1}[2 \hat{a} \rho(t) \hat{a}^{\dagger}-\hat{a}^{\dagger}\hat{a}\rho(t)-\rho(t) \hat{a}^{\dagger}\hat{a}] \nonumber\\ &&+
  \Gamma_{1} [2 \hat{b} \rho(t) \hat{b}^{\dagger}-\hat{b}^{\dagger}\hat{b}\rho(t)-\rho(t) \hat{b}^{\dagger}\hat{b}], 
  \end{eqnarray} 
which is identical to the local master equation \eqref{mastereqntwoosc3} obtained in the limit $\bar{n}_{a}=\bar{n}_{b}={\rm 0}$. Thus, for both reservoirs at zero temperature, and under the RWA on the coupled oscillator Hamiltonian, the local and non-local descriptions coincide. This is a result that will be commented on further when we discuss  in Section \ref{sec:SteadyState} the steady state solutions to the local and non-local master equations.

\subsection{The characteristic function}
 
From the master equation \eqref{disspfullappdxnewoldnayi} we obtain a  partial differential equation for the two-mode quantum characteristic function,
\begin{equation}\label{chievnaycrt}
\frac{\partial }{\partial t} \chi
=z^{T}  {\mbox{\bf M}_1} z \chi+z^{T}  {\mbox{\bf N}_1} \nabla \chi,
\end{equation}
where
$\nabla= (\frac{\partial}{ \partial \kappa_{a}}, \frac{\partial }{ \partial \kappa_{a}^{*}}, \frac{\partial}{ \partial \kappa_{b}}, \frac{\partial }{ \partial \kappa_{b}^{*}})^{T}$
and
\begin{widetext}
\begin{eqnarray}
{\mbox{\bf N}_{1}}&=&\left( \begin{array}{cccc}
   i\omega+\Gamma_{2}-\Gamma_{1} & 0 & \Gamma_{5}-\Gamma_{4}+i \kappa &-i\lambda\\
     0 & -i\omega+\Gamma_{2}-\Gamma_{1}  & i\lambda &-i \kappa +\Gamma_{5}-\Gamma_{4}  \\
      i \kappa +\Gamma_{5}-\Gamma_{4} & -i\lambda & i\omega+\Gamma_{2}-\Gamma_{1}&0 \\
       i\lambda & -i \kappa  +\Gamma_{5}-\Gamma_{4}&0 &-i\omega+\Gamma_{2}-\Gamma_{1}\\
  \end{array} \right )\\
 &&\nonumber\\
  &&\nonumber\\
{\mbox{\bf M}_{1}}&=&\left( \begin{array}{cccc}
  -\Gamma_{3} & -\Gamma_{2}  & i \lambda/2-\Gamma_{6} & -\Gamma_{5} \\
     -\Gamma_{2}  & -\Gamma_{3} & -\Gamma_{5}&-i \lambda/2-\Gamma_{6}\\
    i \lambda/2-\Gamma_{6}& -\Gamma_{5} & -\Gamma_{3}&-\Gamma_{2} \\
       -\Gamma_{5} & -i \lambda/2-\Gamma_{6}&-\Gamma_{2}  &-\Gamma_{3} \\
  \end{array} \right ).
  \end{eqnarray}
  \end{widetext}
Using the numerical solution of equations \eqref{chievnay} and \eqref{chievnaycrt}, equivalently, equations \eqref{mastereqntwoosc3} and \eqref{disspfullappdxnewoldnayi}, we can compare the time evolution of the state of the two coupled oscillators initially prepared in Gaussian states when it evolves according to the master equation \eqref{mastereqntwoosc3} and when it evolves according to \eqref{disspfullappdxnewoldnayi}. This is the subject of the next section.

\section{Time evolution}
\label{sec:resltsapp}
By numerically solving the master equations obtained in sections \ref{sec:pheno} and \ref{sec:mastersolexct}, we can now easily compare the results of the two approaches. We are interested in studying the time evolution of coupled harmonic oscillators initially prepared in Gaussian states. The state of the coupled oscillators can therefore be fully characterized in terms of the covariance matrix. 

\subsection{Oscillator excitation}

\begin{figure}
\centering
\subfigure[]{\label{figent1}
\includegraphics[width=0.42\textwidth]{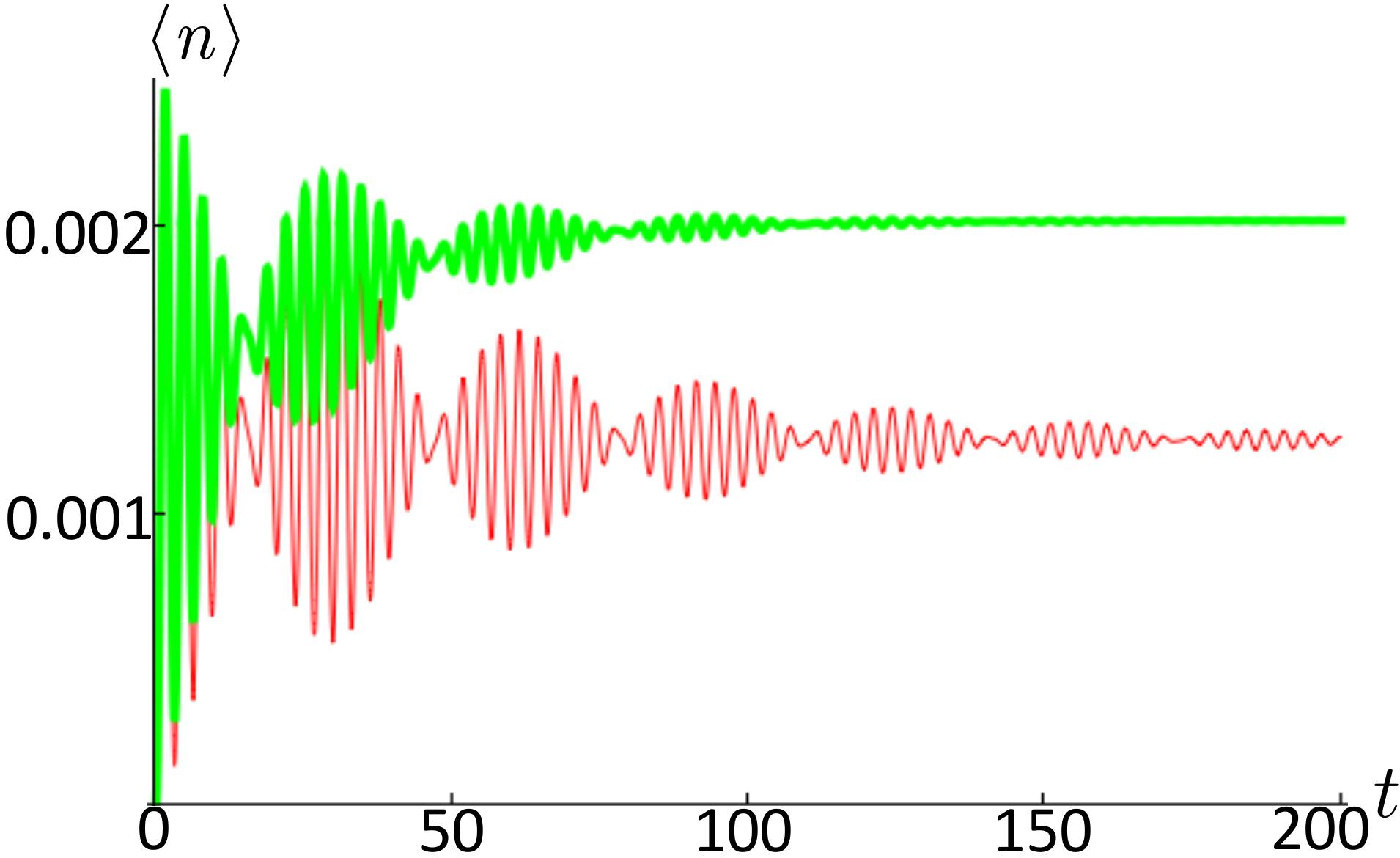}
}
\subfigure[]{\label{figent2}
\includegraphics[width=0.42\textwidth]{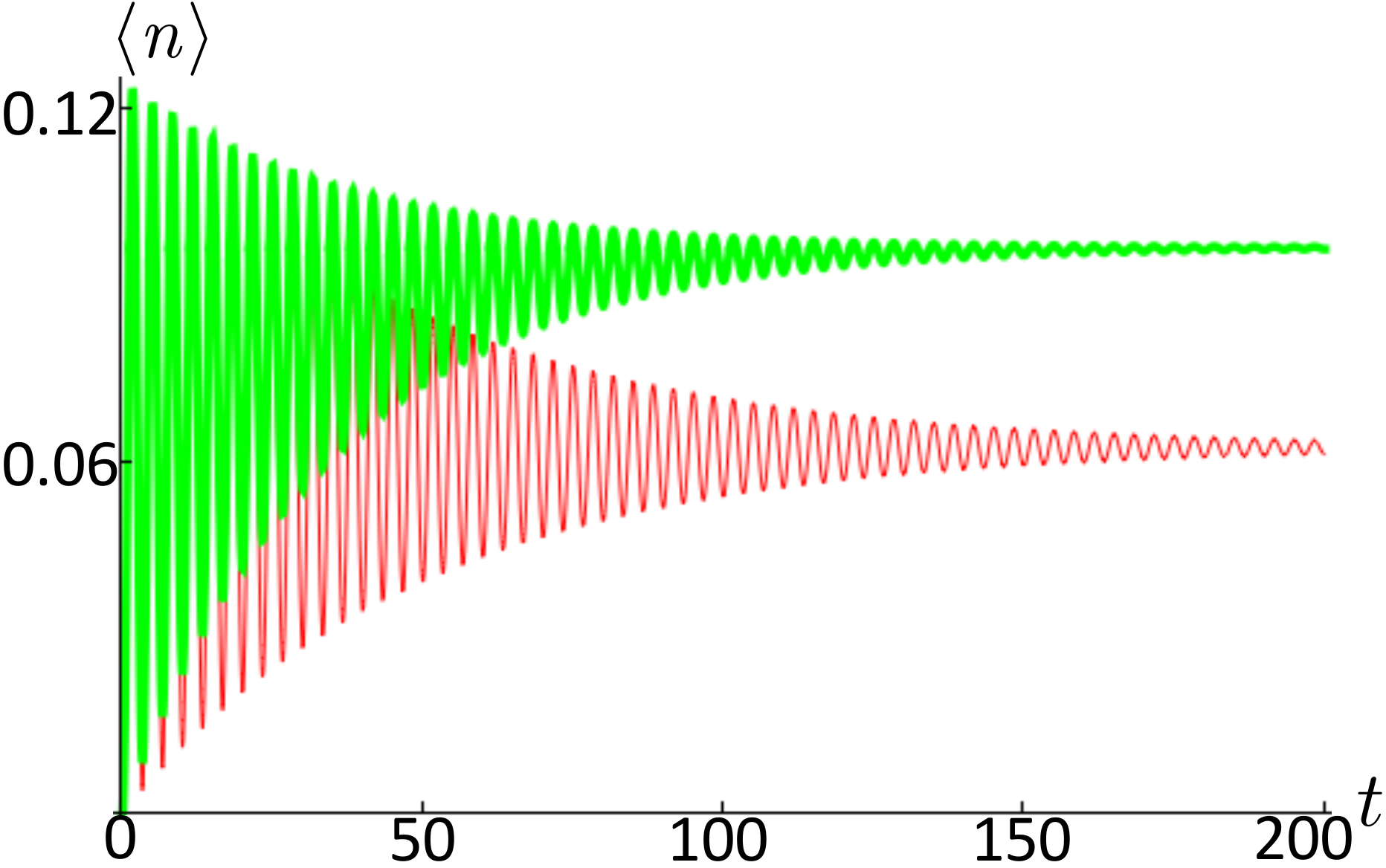}
}
\caption{(Color online) Average number of excitation quanta $n =\langle \hat{a}^{\dagger}({\it t})\hat{a}({\it t})\rangle=\langle \hat{b}^{\dagger}({\it t})\hat{b}({\it t})\rangle$ for each 
oscillator, each of them interacting with an environment in an identical way,
calculated using the master equations \eqref{mastereqntwoosc3} (red, solid) and \eqref{disspfullappdxnewoldnayi} (green, thick solid), plotted as a function of time. Each oscillator is initially in a vacuum state, and $\Gamma_{a}=\Gamma_{b}=\omega$/100. In $(a)$, $\kappa=\lambda=\omega/20$, and in $(b)$~$\lambda=\omega/3$ and $\kappa=0$. Time is in units of $1/\omega$.}
\label{figentnumber} 
\end{figure}

Figures~\ref{figent1} and \ref{figent2} show the average number of excitations for each coupled oscillator evolving according to the local master equation \eqref{mastereqntwoosc3} and the nonlocal master equation \eqref{disspfullappdxnewoldnayi}. One can clearly see that the dissipative dynamics is different depending on which master equation and model for dissipation is used. As will be discussed later, the difference between the two approaches will become even stronger when one looks at the steady-state solutions of the two master equations obtained through the above two approaches.

\subsection{Oscillator correlation}

\begin{figure}
\centering
\subfigure[]{\label{figent3}
\includegraphics[width=0.42\textwidth]{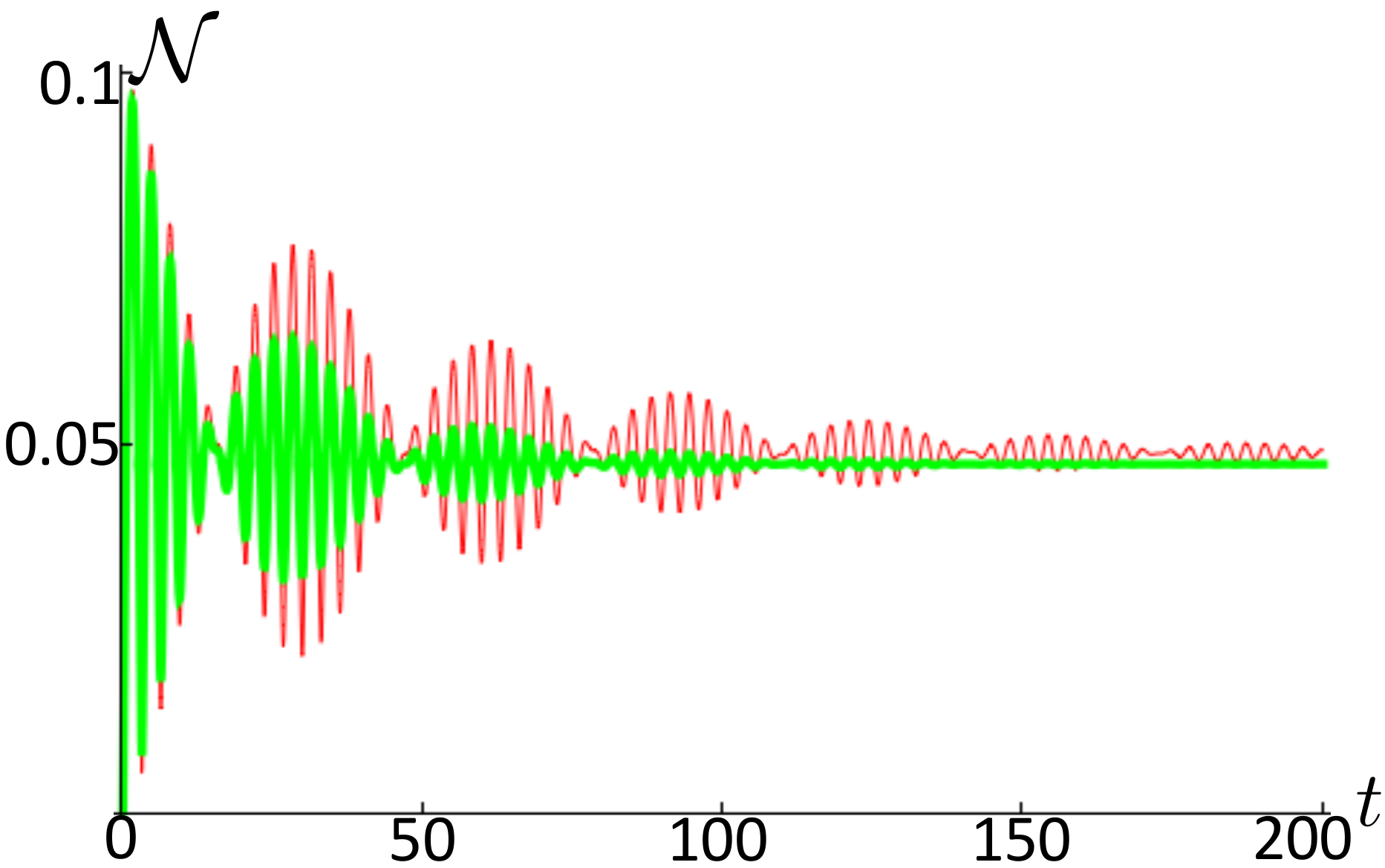}
}
\subfigure[]{\label{figent4}
\includegraphics[width=0.42\textwidth]{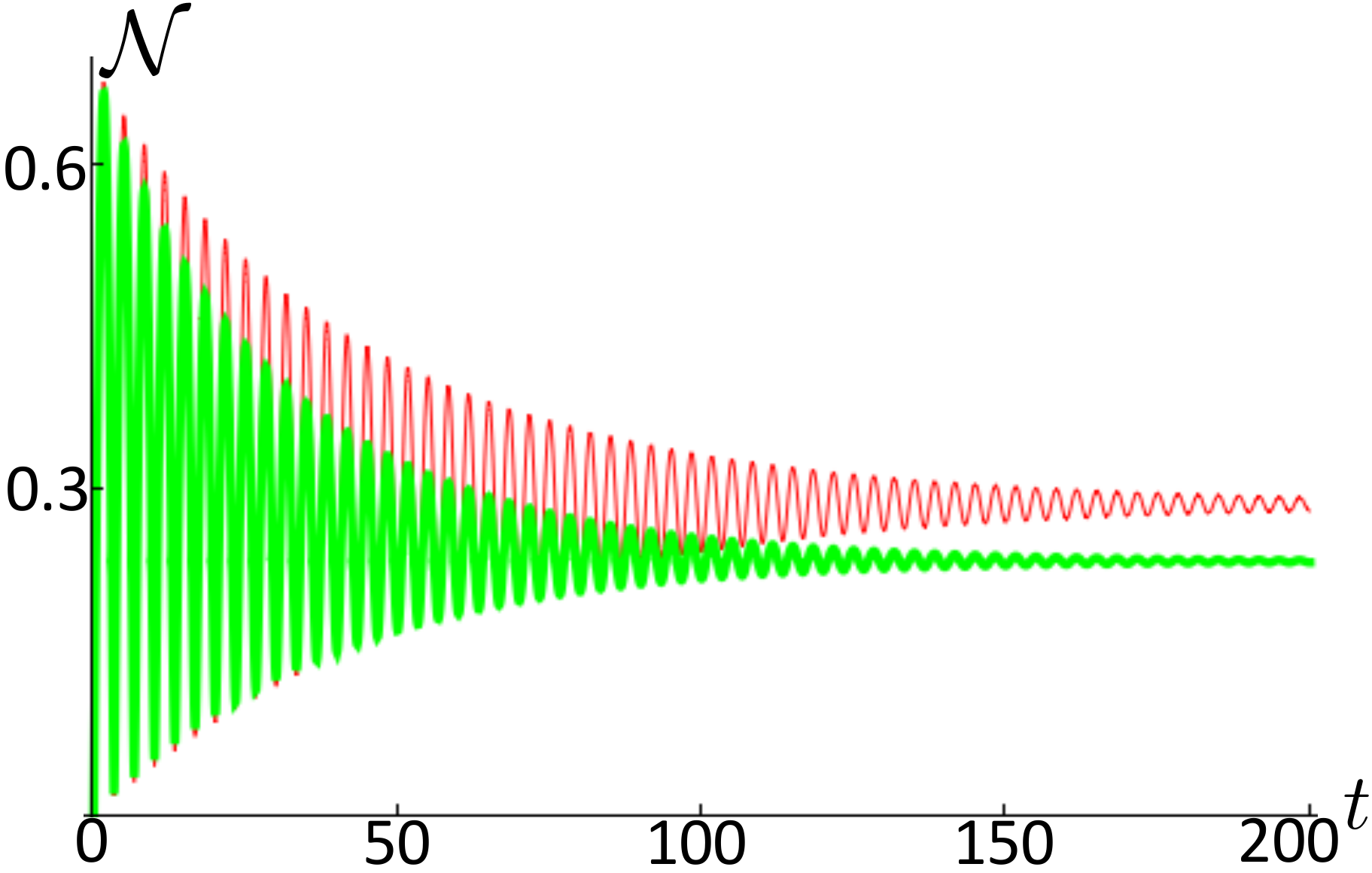}
}
\caption{(Color online)  The logarithmic negativity plotted as a function of time, calculated using 
numerical solutions of the master equations \eqref{mastereqntwoosc3} (red, solid) and \eqref{disspfullappdxnewoldnayi} (green, thick solid). Each oscillator is initially in a vacuum state, and $\Gamma_{a}=\Gamma_{b}=\omega$/100. In $(a)$, $\kappa=\lambda=\omega/20$, and in $(b)$~$\lambda=\omega/3$ and $\kappa=0$. Time is in units of $1/\omega$.}
\label{figentnumber1} 
\end{figure}

To quantify the quantum correlations between the two coupled oscillators, we investigate the entanglement between the oscillators initially prepared in Gaussian separable states. For the case of two-mode Gaussian states, the covariance matrix ${\mbox{\bf V}}$ is a $4 \times 4$ symmetric matrix with $V_{ij}=(\langle R_{i} R_{j}+R_{j} R_{i} \rangle )/2$ where $i,j \in \{a,b\}$ and $R^{T}=(\hat{q}_{a}, \hat{p}_{a},\hat{q}_{b},\hat{p}_{b})$. Here $\hat{q}_{i}$ and $\hat{p}_{i}$ are the position and momentum quadratures of the $i$th oscillator. To characterize the entanglement dynamics we use the logarithmic negativity, which is an entanglement monotone and relatively easy to compute. For a two-mode Gaussian continuous-variable state with covariance matrix $\bf V$, the logarithmic negativity is obtained as $ \mathcal N= \rm{Max}[0,-\rm{ln}(2 \nu_{-})]$ \cite{gera}, where $\nu_{-}$ is the smallest of the symplectic eigenvalues of the covariance matrix, given by $
 \nu_{-}= \sqrt{\sigma/2-\sqrt{(\sigma^{2}-4 \rm{Det} \bf{V})}/\rm{2}}$. Here
\begin{eqnarray}
\sigma&=&\rm{Det} \bf{A_{1}}+\rm{Det} \bf{B_{1}}-\rm{2Det} \bf{C_{1}}\\ \bf V&=&
\left(
\begin{array}{cc}
  \bf A_{1} &\bf C_{1}\\
 \bf C_{1}^{T} & \bf B_{1}
\end{array}
\right),
\end{eqnarray} 
where $\bf {A_{1}}(\bf {B_{1}})$ accounts for the local variances of mode $a(b)$ and $\bf C_{1} $ for the inter-mode correlations. Using the numerical solutions of the partial differential equations \eqref{chievnay} and \eqref{chievnaycrt} we compute the logarithmic negativity, shown in Fig.~\ref{figent3} and Fig.~\ref{figent4}. As can be seen from these figures, the two different approaches for modeling the system-reservoir interactions, discussed in sections \ref{sec:pheno} and \ref{sec:mastersolexct}, yield quantitatively very different results as far as quantum correlations between the two oscillators are concerned.

\subsection{Fidelity}

The difference in the dynamics for the two approaches can be further illustrated by computing the quantum fidelity between the time-evolved states of the two oscillators. In general, finding the fidelity between two quantum states is difficult, but for Gaussian states it is possible to arrive at a closed-form expression for the quantum fidelity in terms of the covariance matrix. We trace over the state of one of the oscillators, and compute the fidelity between the two different single-oscillator states resulting from the numerical solutions of equations \eqref{chievnay} and \eqref{chievnaycrt}. 

The one-mode quantum characteristic function $\chi(\kappa_{a})$ can be deduced from the two-mode quantum characteristic function $\chi(\kappa_{a},\kappa_{b})$ through the identity $\chi(\kappa_{a},{\it t})=\chi(\kappa_{a},\kappa_{b}={\rm 0},{\it t})$. In this way a one-mode Gaussian state of the pair of oscillators can be defined, which is used for calculating the corresponding fidelity. 

\begin{figure}
\subfigure[]{\label{figent5}
\includegraphics[width=0.42\textwidth]{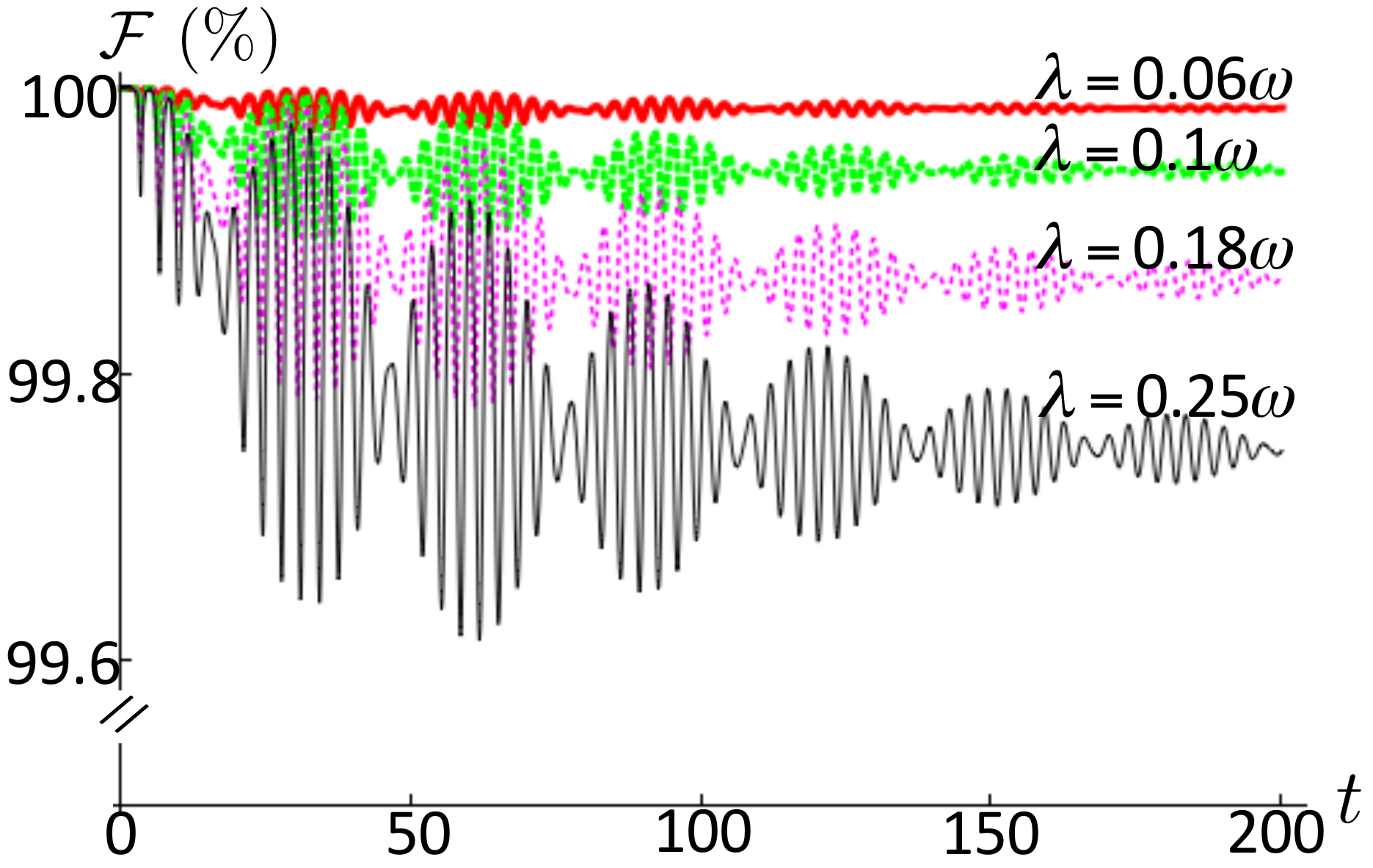}
}
\subfigure[]{\label{figent6}
\includegraphics[width=0.42\textwidth]{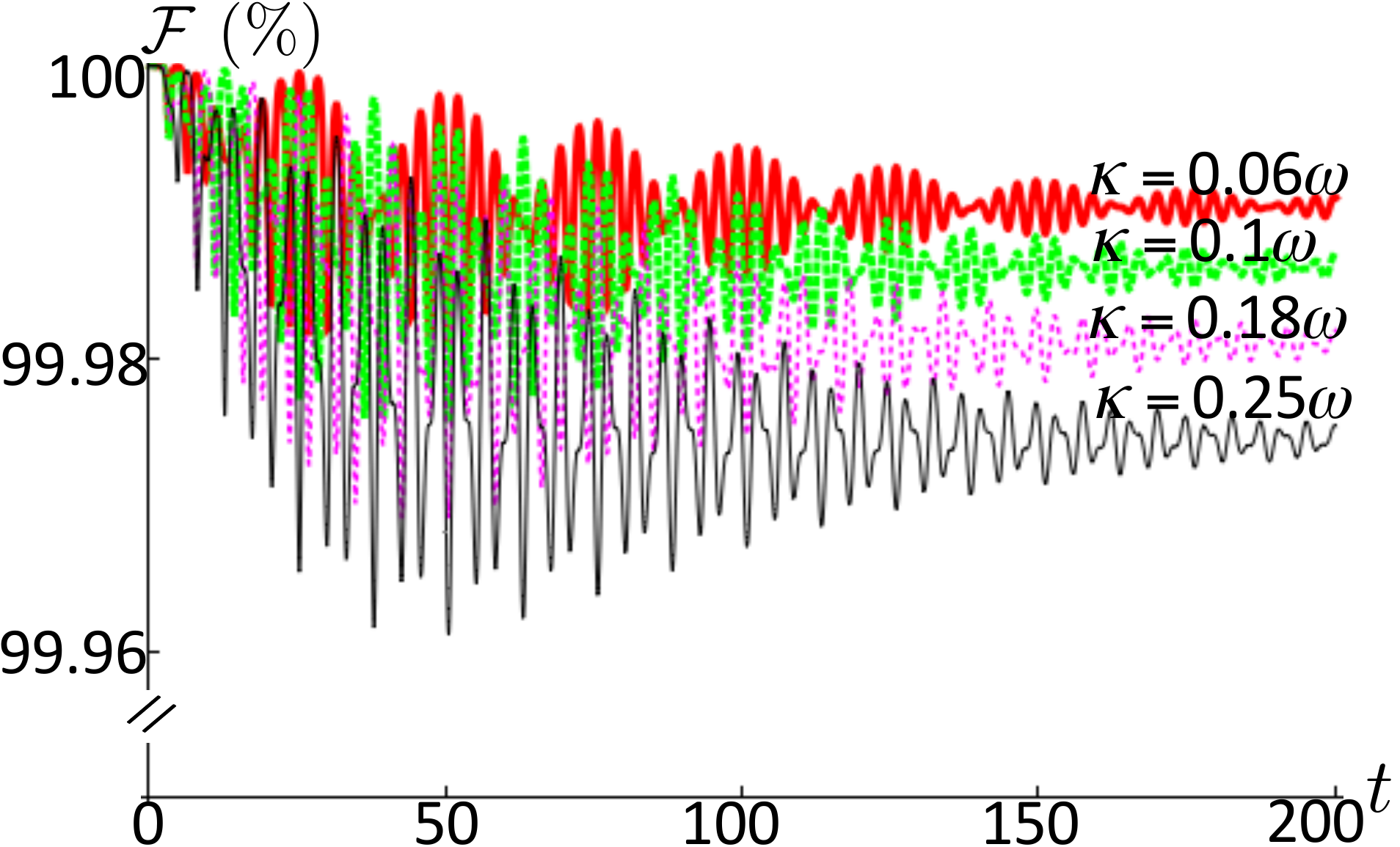}
}
\caption{(Color online) Time dependence of the quantum fidelity between the two one-mode states of each oscillator computed from the numerical solutions of equations \eqref{mastereqntwoosc3} and \eqref{disspfullappdxnewoldnayi} for $\Gamma_{a}=\Gamma_{b}=\omega/100$, when $(a)$~ $\kappa=\omega/20$ and, $(b)$~ $\lambda = \omega/20$. Each oscillator is initially in the vacuum state, and time is in units of $1/\omega$.}
\label{figentnumber2} 
\end{figure}

The quantum fidelity between two one-mode Gaussian states can be computed from
\begin{equation}
\mathcal F= \frac{2}{(\sqrt{{\rm Det}[{\bf A_{1}}+{\bf A_{2}}]+\mathcal P}-\sqrt{ \mathcal P})}
\end{equation}
where 
\begin{equation} 
\mathcal P=({\rm Det}[\bf A_{1}]-1)({\rm Det}[{\bf A_{2}}]-1),
\end{equation}
and where ${\bf A_{i}}$ is the 2$\times$2 covariance matrix corresponding to the {\it i}:th mode \cite{hscut,ghspar}. The time evolution of the fidelity between the solutions of sections \ref{sec:pheno} and \ref{sec:mastersolexct} is shown in Fig.~\ref{figent5} and Fig.~\ref{figent6}, where the initial state was chosen to be the ground state of each oscillator. As can be seen from these figures, when the inter-mode coupling strength between the oscillators increases, the fidelity between the time-evolved one-mode Gaussian states of each oscillator obtained through the solution of master equations \eqref{mastereqntwoosc3} and \eqref{disspfullappdxnewoldnayi} decreases. Thus it is evident that if the oscillators are strongly coupled then the solution of the master equation \eqref{mastereqntwoosc3} starts to disagree with the solution of the master equation \eqref{disspfullappdxnewoldnayi}. Nonetheless the fidelity between the two solutions stays much above 99 $\%$ for a wide range of coupling strengths $\kappa, \lambda$. It should be noted from Fig.~\ref{figent5} and ~\ref{figent6} that the mismatch between the solutions of the master equations \eqref{mastereqntwoosc3} and \eqref{disspfullappdxnewoldnayi} becomes more prominent if the two-mode squeezing interaction strength $\lambda$ increases. The above observations from Fig.~\ref{figentnumber2} remain qualitatively unchanged 
even when the two oscillators are initialized in a separable squeezed state.

\subsection{Steady state\label{sec:SteadyState}}
 
 Recall the original observation by Walls~\cite{wallsp} that the validity of the local Lindblad master equation is open to question on the basis that it fails to derive the expected thermal equilibrium density operator for the system. We can investigate the issue here by first returning to the local master equation \eqref{mastereqntwoosc3} for both reservoirs at zero temperature, where $\bar{n}_{a}=\bar{n}_{b}={\rm 0}$
\begin{equation}\label{disspfullappdxnewdif}
	\dot{\rho}=-i\left[H_\text{sys},\rho\right]+\Gamma_a\mathcal{L}_a\rho+\Gamma_b\mathcal{L}_b\rho
\end{equation}
and where $\mathcal{L}_{l}\rho=\hat{l}\rho\, \hat{l}^\dagger-\frac{1}{2}\left\{ \hat{l}^\dagger\hat{l},\rho\right\}$. Now it is a simple matter to verify that the steady state of the non-local master equation \eqref{disspfullappdxnewold} is $\rho_{ss}$=$|0\rangle_{ll} \langle 0| \otimes |0\rangle_{mm} \langle 0 |$, while the steady state of the local master equation \eqref{disspfullappdxnewdif} is $\rho_{ss}=|0\rangle_{aa}\langle 0|\otimes|0\rangle_{bb}\langle 0|$. The point to note is that the vacuum states of the non-local $\hat{l},\hat{m}$ oscillators are not the same as those of the local $\hat{a},\hat{b}$ oscillators, as can be easily demonstrated by using the relation between $\hat{a}$ and $\hat{l},\hat{m}$ and their Hermitean conjugates, readily obtainable from \eqref{efab} and \eqref{secappndx3}, to show that
\begin{equation}
	\hat{a}|0\rangle_l|0\rangle_m
	=-\frac{1}{\!\sqrt{2}}\left(\beta_1|1\rangle_l|0\rangle_m+\beta_2|0\rangle_l|1\rangle_m\right)
\end{equation}
(and similarly for $\hat{b}|0\rangle_l|0\rangle_m$) which only vanishes if $\beta_1=\beta_2=0$ which implies $\lambda=0$, i.e., the full RWA Hamiltonian, a not unexpected result since for $\kappa \ne 0$ and $\lambda=0$, and in the zero temperature limit of $\bar{n}_{a}=\bar{n}_{b}={\rm 0}$, the non-local master equation \eqref{disspfullappdxnewold} is identical to the master equation \eqref{disspfullappdxnewdif}, as already indicated in \eqref{mastreqnewjuly}. The steady state is then the separable trivial ground state of each operator.

In general, it is the non-local steady state which is the limit for zero temperature of the canonical density operator \eqref{canonical}: the ground state of the generalised Hamiltonian \eqref{eqn2ham}. This result extends Walls's early result to the case of two coupled bosonic modes interacting under this generalized Hamiltonian, at least for a zero temperature reservoir. We have further shown that this result may be of great significance in understanding the entanglement properties of the ground state of coupled harmonic oscillators.

\begin{figure}
\subfigure[]{\label{figent7}
\includegraphics[width=0.42\textwidth]{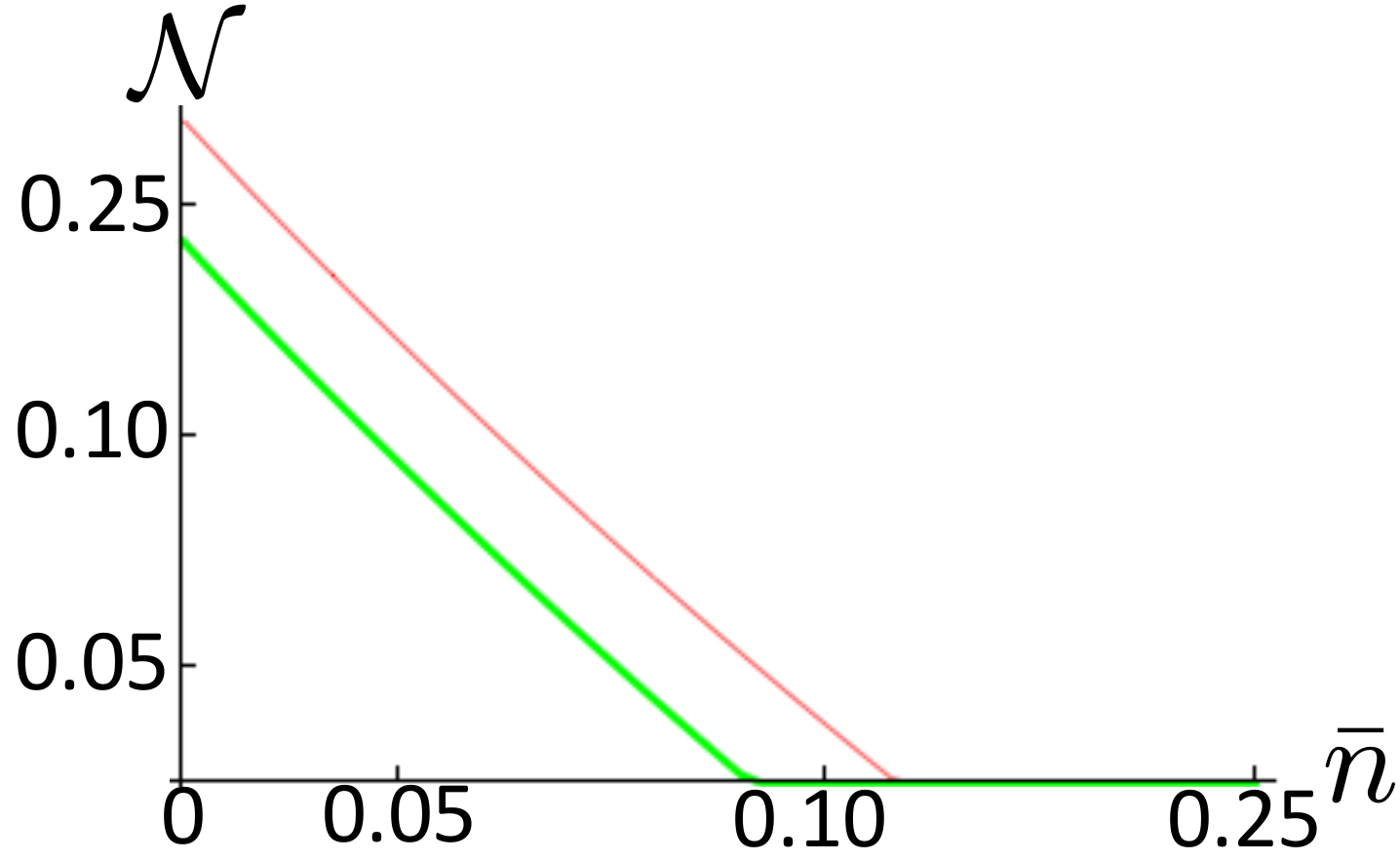}
}
\subfigure[]{\label{figent8}
\includegraphics[width=0.42\textwidth]{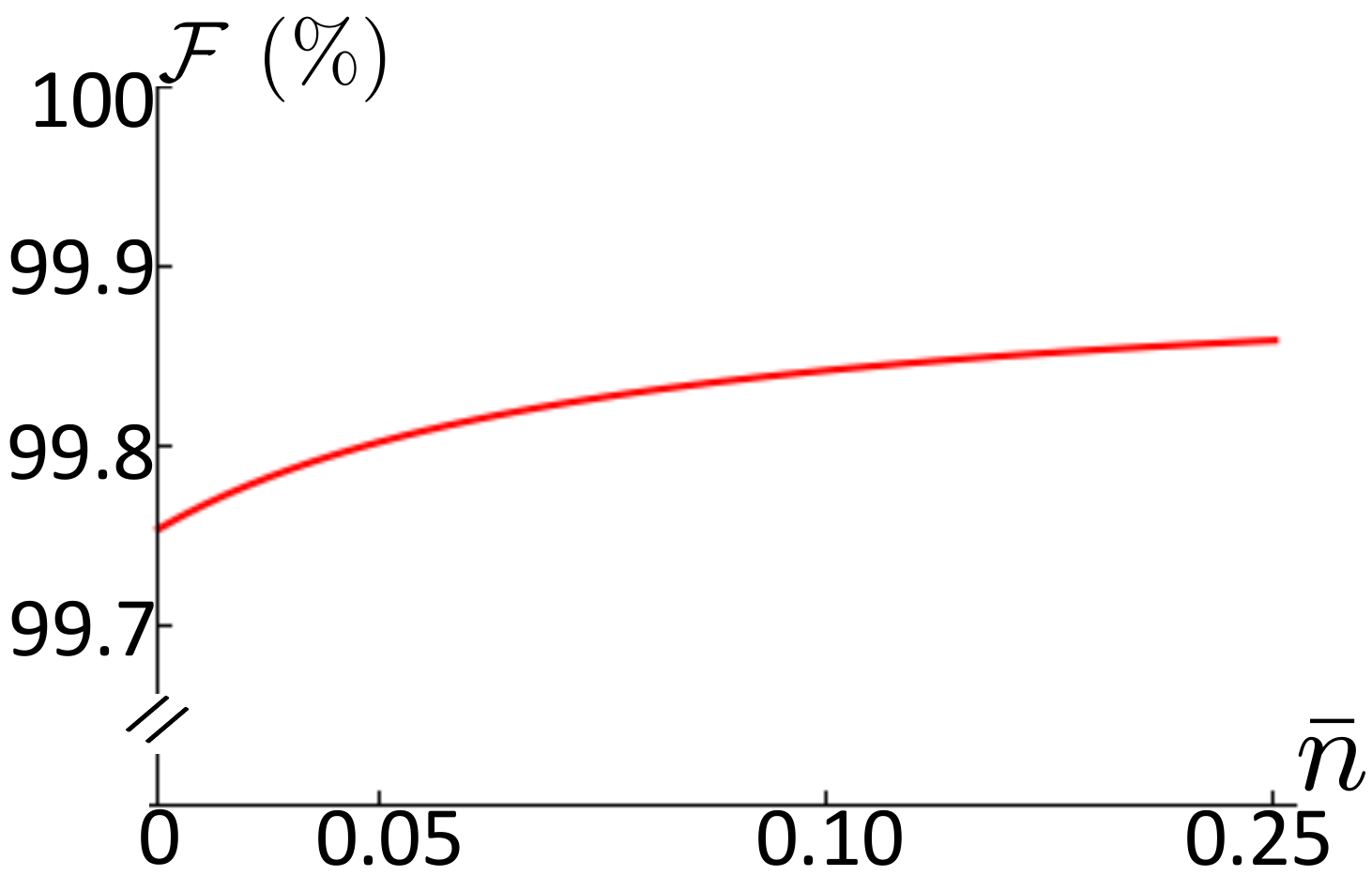}
}
\caption{(Color online) Steady state value of $(a)$ the logarithmic negativity calculated using the local Lindblad-type dissipation approach (red, solid) and the bath-induced dissipation approach (green, thick solid), and $(b)$ the quantum fidelity between the two one-mode states of each oscillator obtained  using the approaches of sections \ref{sec:pheno} and \ref{sec:mastersolexct}, plotted as a function of $\bar{n}$. Other physical parameters are chosen such that $\Gamma_{a}=\Gamma_{b}=\omega$/100, $\lambda=\omega/3$ and $\kappa=0$.}
\label{figentfidtemp} 
\end{figure}

Before concluding this section we will briefly consider a physical scenario where  each coupled oscillator is in contact with 
an identical heat bath with non-zero average thermal occupancy ($\bar{n}_{a}=\bar{n}_{b}= \bar{n}\neq 0$). It is a straightforward exercise  to extend the master equation \eqref{disspfullappdxnewold} to include the thermal fluctuations of the heat bath for each oscillator. We, however,  do not explicitly detail this calculation and only report the results here.  Fig.~\ref{figentfidtemp}  shows the  steady state  value of  the logarithmic negativity, and  the quantum fidelity between the two one-mode states of each oscillator obtained  using the approaches of sections \ref{sec:pheno} and \ref{sec:mastersolexct}, plotted as a function of $\bar{n}$. As can be seen from Fig.~\ref{figent7}, a critical value of thermal noise destroys the pairwise entanglement between the oscillators. Also evident is the feature that the phenomenological modeling of dissipation, as compared to the bath-induced dissipation approach of  section \ref{sec:mastersolexct}, overestimates the magnitude of steady state  logarithmic negativity. A noteworthy feature  of Fig.~\ref{figent8} is that with increasing $\bar{n}$, the quantum fidelity between the two one-mode states of each oscillator, obtained using the approaches of sections \ref{sec:pheno} and \ref{sec:mastersolexct}, improves further.
However, it is worth mentioning that even in a regime when the coupled oscillators are in a separable state ($\bar{n} \sim 0.12$), the quantum fidelity between the two one-mode states of each oscillator, obtained  using the approaches of sections \ref{sec:pheno} and \ref{sec:mastersolexct}, does not reach unity.

\section{Discussion and summary}
\label{sec:concld}  

To summarize, we have investigated Markovian master equations for two  harmonic oscillators, coupled through a general Hamiltonian \eqref{eqn2ham}. We especially considered the regime where the oscillators are strongly coupled to each other, in which case the RWA cannot be applied to an oscillator-oscillator coupling of the form in Eq.\ \eqref{eq:sysham}. We compared two situations. First, a case where the dissipation of each oscillator was modeled with local Lindblad terms, added ``phenomenologically" for each individual oscillator. This situation was then compared to the case where each oscillator is coupled to a bath of harmonic oscillators, resulting in a master equation \eqref{disspfullappdxnewoldnayi} of Lindblad form, but where the Lindblad terms are not local in terms of the individual oscillator modes. Specifically, in our derivation of Eq.\ \eqref{disspfullappdxnewoldnayi}, the RWA for the system-environment interaction is made at the level of mode operators for the eigenmodes $\hat l$, $\hat m$ of the total oscillator-oscillator Hamiltonian including the coupling between the oscillators. The master equation with local Lindblad operators in Eq.\ \eqref{mastereqntwoosc3} would result if one instead drops terms involving the individual oscillator modes $\hat a$ and $\hat b$, of the form $\hat a \hat c, \hat a^\dagger \hat c^\dagger, \hat b \hat d$ and $\hat b^\dagger \hat d^\dagger$.  However, this is not a correct application of the RWA, as it does not correctly identify, and then remove, the rapidly oscillating terms.
 
As we have shown, the difference between the two approaches will in fact result in different steady state solutions, which may give rise to non-trivial differences in ground state properties, especially with regards to non-classical correlations such as entanglement. Modeling system-environment interaction through local Lindblad operators may be valid if the inter-mode couplings are weak. However, in a system of strongly coupled bosonic modes, modeling dissipation through local damping of each individual mode may become questionable and may give rise to dubious results. 

We like to briefly comment that there are equivalent  ways to {\it exactly} solve the open dynamics of two coupled harmonic oscillators. This includes the Heisenberg-Langevin equation approach and the Feynman and Vernon path integral approach for open quantum systems \cite{dissiptvebook2}. In this direction, previous works have addressed  coupled harmonic oscillators interacting with common or independent  baths with arbitrary spectra \cite{noteref}. In this work, however, we have worked in the Schr\"odinger picture and have derived a master equation for a pair of harmonic oscillators  coupled under the generalized Hamiltonian \eqref{eqn2ham}. Provided the Born-Markov and secular approximations  hold and the two oscillators  interact with their independent heat baths (with flat spectra), the master equation \eqref{disspfullappdxnewoldnayi} is {\it exact} in modeling the dissipative dynamics of the two oscillators. We have subsequently converted the master equation \eqref{disspfullappdxnewoldnayi}  into a Fokker Planck equation and have solved it numerically. Therefore, all our results are exact within the ambit of above approximations.


\acknowledgments{We gratefully acknowledge fruitful discussions with Michael Hall and Steve Barnett.}

\end{document}